\documentclass[pra,twocolumn,groupedaddress,showpacs]{revtex4}
\usepackage{graphics}
\usepackage{enumerate}
\usepackage{amsmath}
\begin{document}
\newcommand{\op}{\boldsymbol}
\bibliographystyle{apsrev}

\title{Master Key Secured Quantum Key Distribution}

\author{Tabish Qureshi}
\email{tabish@ctp-jamia.res.in}
\affiliation{Centre for Theoretical Physics, Jamia Millia Islamia, New Delhi, India.}
\author{Tabish Shibli}
\email{tabishy2k7@gmail.com}
\affiliation{SGTB Khalsa College, New Delhi, India.}
\author{Aditi Sheel}
\email{mail2adt@gmail.com}
\affiliation{Department of Physics, Jamia Millia Islamia, New Delhi-110025,
India.}


\begin{abstract}
A new scheme of Quantum Key Distribution is proposed using three entangled
particles in a GHZ state. Alice holds a 3-particle source and sends two particles to Bob,
keeping one with herself. Bob uses one particle to generate a secure key,
and the other to generate a {\em master-key}. This scheme should prove to
be harder to break in non-ideal situations
as compared to the standard protocols BB84 and Eckert. The scheme
uses the concept of Quantum Disentanglement Eraser. Extension to multi-partite
scheme has also been investigated.
\end{abstract}

\pacs{03.67.Dd ; 03.65.Ud}
\maketitle

Crypting messages for secret communication is a very old problem. The
so-called Vernam Cipher \cite{vernam} or one-time pad, is a method
which is believed
to be the most secure, with the caveat that it is based on a shared key
which can only be used once. This led to people exploring the possibility
of remotely sharing a new secret key in a secure way.
Quantum key distribution (QKD) allows two parties, conventionally called
Alice and Bob, to generate a common
string of secret bits, the secret key, in the presence
of an eavesdropper, usually called Eve \cite{qkd}. The key so generated,
may be used for crypting messages using Vernam Cipher.
The pioneering protocol for QKD was given by
Bennet and Brassard in 1984, in a conference in Bangalore \cite{bb84}.
Later another equivalent protocol was given by Eckert utilizing properties
of entangled states \cite{eckert}.
In principle, QKD is hundred percent secure, the proof being provided
by the laws of quantum mechanics \cite{shor}.  However,
real-life implementations of QKD have various issues which make them
deviate from the assumptions in idealized models.
By exploiting security loopholes in practical realizations, notably
imperfections in the detectors, various
attacks have been successfully demonstrated against commercial QKD
systems \cite{qkdbreak,qkdbreak1}.

Here we introduce a new QKD method using three entangled particles. This
method introduces an additional element in the standard key distribution
protocols, to make it harder to break in non-ideal situations.

\section{BB84 and Eckert Protocols}

The basic quantum key distribution protocol of BB84 \cite{bb84} or Eckert
\cite{eckert} is as follows. 

\begin{enumerate}
\item An entangled spin-1/2 particle source produces
a sequence of particles pairs, {\em in a singlet state}, one going to Alice,
and one to Bob. 
\item Bob measures the incoming particles' spin states by randomly choosing a 
measurement of either the x-component of the spin or the z-component,
with equal probability.
\item Bob publicly tells Alice which bases he used for each particle he received
(but, of course not the result of his measurement).
\item Alice publicly tells Bob which bases she used to measure her particles.
\item Alice and Bob keep only the data from those measurements for which their
bases are the same, discarding all the rest.
\item This data is interpreted as a binary sequence according to the coding
scheme $|+\rangle_x$ = 1, $|-\rangle_x$ = 0, $|+\rangle_z$ = 1,
$|-\rangle_z$ = 0 for Alice, and
$|+\rangle_x$ = 0, $|-\rangle_x$ = 1, $|+\rangle_z$ = 0,
$|-\rangle_z$ = 1 for Bob.
\item Alice announces the results of a small subset of her measurements. 
Bob checks if he has identical results. Any discrepency here indicates
a possible evesdropping attempt.
\item If there is no discrepancy, the rest of the binary sequence is treated
as the new key, and is identical for both Alice and Bob.
\end{enumerate}

If the entangled-particle source is held by Alice, and only one particle
travels to Bob and the other remains with Alice, the protocol is essentially
BB84. She could replace it by a source producing single particles, each of
which she measures before forwarding it to Bob. The consequences will be
identical to those described above.

\section{Three particle entanglement}

Let us consider the following 3-particle entangled state, known
as the GHZ state \cite{ghz}
\begin{equation}
|\psi\rangle=\frac{1}{\sqrt{2}}(|\uparrow\rangle_{1}|\uparrow\rangle_{2}
|\uparrow\rangle_{3}+|\downarrow\rangle_{1}|\downarrow\rangle_{2}
|\downarrow\rangle_{3}),
\label{entstate1}
\end{equation}
where the states $|\uparrow\rangle_{i}, |\downarrow\rangle_{i}$ are 
eigenstates of the operator $\op{\sigma}_{iz}$.
and, let us also consider the following transformation in basis, 
\begin{eqnarray}
|\uparrow\rangle_{i}=\frac{1}{\sqrt{2}}(|+\rangle_{i}+|-\rangle_{i}),~~~~
|\downarrow\rangle_{i}=\frac{1}{\sqrt{2}}(|+\rangle_{i}-|-\rangle_{i})
\end{eqnarray}
for i=1,2 and 3. If we just look at the subspaces of particles 1 and 2, their
state is not a pure entangled state, but a mixed state, as can be seen
by writing the density matrix for (\ref{entstate1}) and tracing over the
states $|\uparrow\rangle_{3}, |\downarrow\rangle_{3}$. Here, the results of
measurement of $\op{\sigma}_{1z}$ and $\op{\sigma}_{2z}$ will be correlated, but
results of measurements of $\op{\sigma}_{1x}$ and $\op{\sigma}_{2x}$ will not be
correlated.

Writing the states of particle 3 in terms of the eigenstates of
$\op{\sigma}_{3x}$, (\ref{entstate1}) can be written as
\begin{eqnarray}
|\psi\rangle&=&\frac{1}{2}(|\uparrow\rangle_{1}|\uparrow\rangle_{2}
+|\downarrow\rangle_{1}|\downarrow\rangle_{2})|+\rangle_{3}\nonumber\\
&&+\frac{1}{2}\left(|\uparrow\rangle_{1}|\uparrow\rangle_{2}
-|\downarrow\rangle_{1}|\downarrow\rangle_{2}\right)|-\rangle_{3}.
\label{entstate2}
\end{eqnarray}
As we have not changed the state, measurements on particle 1 and 2 will not
show any quantum correlations. However, if one also makes a measurement of
$\op{\sigma}_{3x}$, and picks out only those results of measurement of
particle 1 and 2, for which particle 3 yields $|+\rangle_{3}$,
particle 1 and 2 will show quantum correlation. Particles 1 and 2, which
appeared to be disentangled in state (\ref{entstate1}), are now
entangled. One can say that a measurement of $\op{\sigma}_{3x}$ has {\em erased}
the disentaglement between particle 1 and 2. Correlating the measurements of
particles 1 and 2 with $|-\rangle_{3}$ will also lead to an
entanglement of 1 and 2, but of a slightly different form. This concept of
{\em quantum disentanglement eraser} was introduced by Garisto and Hardy
\cite{hardy}.

As one can see, measurement of particle 3, in a particular basis, has the
potential to control the nature of entanglement of particles 1 and 2. We use
this feature to construct a new 3-particle protocol for QKD. The GHZ state
has been used before to construct QKD protocol for sharing a secure key
between three parties \cite{khym}. However, we are only interested in two-party
key-sharing.

\section{Master-Key Secured QKD (MKS-QKD)}

In the following we propose a key distribution scheme where Alice 
holds a 3-particle source which generates a sequence of particle trios in
a GHZ state given by (\ref{entstate1}). She sends particle 2 and 3 to Bob and
keeps particle 1 with herself. Bob calls (say) particle 3 as master channel and
the particle 2 as secure channel. He measures 
$\op{\sigma}_{3x}$ on the master channel
so that he gets either $|+\rangle_{3}$ or $|-\rangle_{3}$. One can see from
(\ref{entstate2}) that if Alice and Bob measure $\op{\sigma}_{1z}$ and
$\op{\sigma}_{2z}$ respectively, their
results will always be correlated. For example, if Alice gets
$|\downarrow\rangle_{1}$ Bob will necessarily get $|\downarrow\rangle_{2}$,
and if Alice gets $|\uparrow\rangle_{1}$ Bob will necessarily get
$|\uparrow\rangle_{2}$, irrespective of the results of the master channel.

Now if Alice and Bob (on particle 2) measure $\op{\sigma}_{ix}$, writing
(\ref{entstate1}) in terms of the eigenstates of $\op{\sigma}_{ix}$ will
make it easier to see what will happen.
\begin{eqnarray}
|\psi\rangle&=&\frac{1}{2}(|+\rangle_1|+\rangle_2
+|-\rangle_{1}|-\rangle_2)~|+\rangle_3\nonumber\\
&&+\frac{1}{2}\left(|+\rangle_1|-\rangle_2
+|-\rangle_1|+\rangle_2\right)~|-\rangle_3.
\label{entstate3}
\end{eqnarray}
It is clear from the above that if the master channel measurement gives 
$|+\rangle_3$ the
measurement results of Alice and Bob (on particle 2) on $\op{\sigma}_{ix}$
will be identical.
On the other hand, if the master channel measurement gives $|-\rangle_3$ the
measurement results of Alice and Bob (on particle 2) on $\op{\sigma}_{ix}$
will be inverted with respect to each other.

On receiving the two particles through
two different channels, Bob randomly decides to use one channel to generate his
secure key, and the other to generate a master-key. The details of the
protocol are as follows.

\begin{enumerate}
\item A 3-particle source is held by Alice which generates a sequence of
3 entangled particles. Particle 1 remains with Alice, while particles
2 and 3 go to Bob through two different channels.
\item Bob randomly chooses one channel to generate his secure key and the
other to generate the master-key. Bob randomly chooses a different channel for
his master-key, for each pair that comes to him.
\item Alice measures the incoming particles' spin states by randomly
choosing a measurement of either the x-component of the spin or the z-component,
with equal probability. Bob does the same for his secure channel.
\item Bob measures the x-component of the spin of particles from his master
channel.
\item Alice and Bob publicly declare which bases they used for the secure
channel, for each particle they received.
\item Alice and Bob keep only the data from those measurements for which their
{\em secure channel} bases are the same, discarding all the rest. 
\item This data is interpreted as a binary sequence according to the coding
scheme $|\uparrow\rangle \rightarrow$ 1, $|\downarrow\rangle \rightarrow$ 0,
$|+\rangle \rightarrow$ 1,
$|-\rangle \rightarrow$ 0 by Alice and Bob.\\
Bob interprets the data of the master channel as
follows: $|+\rangle \rightarrow$ 0, $|-\rangle \rightarrow$ 1, if he measured
x-component in the secure channel;
$|+\rangle \rightarrow$ 0, $|-\rangle \rightarrow$ 0, if he measured
z-component in the secure channel.
Alice and Bob's key doesn't match at this stage.
\item Bob now adds the master-key to his key bit by bit, modulo 2.
\item At this stage, the keys generated by Alice and Bob are identical.
\item In order to check for any evesdropping attempt, Alice announces the
results of a small subset of her measurements. 
Bob checks if he has identical results. Any discrepency here indicates
a possible evesdropping attempt. The rest of the sequence now forms the
usable key.
\end{enumerate}

In the nearly impossible scenario if an evesdropper correctly guesses which
is the master channel for each pair of particles that travels to Bob, he can
perform measurement of $\op{\sigma}_{mx}$, where $m$ is the particle number
which is considered to be the master channel, and can know in advance Bob's
master-key. The security of this key distribution scheme then reduces to
that of the Eckert
or BB84 protocol. However, there is no way an evesdropper can correctly
guess which one is the master channel for every single pair. Evesdropper
measuring $\op{\sigma}_{x}$ on the wrong channel will lead to his attempt
being detected. This feature
introduces an additional complexity in the secure key distribution, and
consequently makes the key sharing more robust against attacks.

\section{Master-Key Controlled QKD (MKC-QKD)}

We now use the concept of disentanglement eraser to construct another
kind of key distribution scheme in which there is a Master who wishes
to {\em control} the key distribution between Alice and Bob.
In this scheme, the key held by the Master has a special position that
without using it Alice and Bob cannot  share a secure key eventhough they
used the Eckert protocol. This is much
like a system in some bank lockers where the bank holds a master-key
without using which the key of an individual client doesn't work.
The protocol for the Master-key controlled quantum key distribution
works as follows.

\begin{enumerate}
\item An 3-particle source is held by the Master which generates a sequence of
3 entangled particles. Particle 1 goes to Alice,
particle 2 to Bob and particle 3 remains with the Master.
\item Alice and Bob measure the incoming particles' spin states by randomly
choosing a measurement of either the x-component of the spin or the z-component,
with equal probability.
\item The Master measures the x-component of the spin of his particle.
\item Bob and publicly declare which bases they used for each particle they
received (but, of course not the result of the measurement).
\item Alice and Bob keep only the data from those measurements for which their
bases are the same, discarding all the rest. The Master also discards the
data for particles for which Alice and Bob's bases do not match.
\item This data is interpreted as a binary sequence according to the coding
scheme $|\uparrow\rangle \rightarrow$ 1, $|\downarrow\rangle \rightarrow$ 0,
$|+\rangle \rightarrow$ 1,
$|-\rangle \rightarrow$ 0 by Alice and Bob. The Master interprets his data as
follows: $|+\rangle \rightarrow$ 0, $|-\rangle \rightarrow$ 1, if Alice and
Bob measured x-component;
$|+\rangle \rightarrow$ 0, $|-\rangle \rightarrow$ 0, if Alice and Bob measured
z-component. All three now
have a key, but Alice and Bob's key doesn't match.
\item The Master announces his key publicly which Bob adds to his key
bit by bit, modulo 2.
\item At this stage, the keys generated by Alice and Bob are identical.
\item In order to check for any evesdropping attempt, Alice announces the
results of a small subset of her measurements. 
Bob checks if he has identical results. Any discrepency here indicates
a possible evesdropping attempt. The rest of the sequence now forms the
usable key.
\end{enumerate}

Using this scheme the Master can effectively delay the sharing of the
key by any length of time. Another possible use of MKC-QKD is that if the
entangled particles are to be provided by a third party, this method
provides a way of authenticating the particle source. Without the publicly
sent master-key, the keys of Alice and Bob will not match.
Although this scheme provides a mechanism which
makes the involement of the Master necessary, it may not provide any 
additional security over the BB84 and Eckert protocols.

\section{Multi-Particle GHZ state}

One might wonder if the n-particle GHZ state
\begin{equation}
|\psi\rangle=\frac{1}{\sqrt{2}}[|\uparrow\rangle_{1}|\uparrow\rangle_{2}
|\uparrow\rangle_{3}\dots|\uparrow\rangle_{n}
+|\downarrow\rangle_{1}|\downarrow\rangle_{2}
|\downarrow\rangle_{3}\dots|\downarrow\rangle_{n}],
\label{ghzn}
\end{equation}
posesses similar properties. In this state too, any two particles are not
entangled, as the two-particle reduced density matrix, after tracing over
rest of the n-2 particles, is a mixed state density matrix. However,
one can show that if one measures the n-2 particles in an appropriate
basis, the entanglement between the two particles can be brought back
by correlating with the measurement results of n-2 particles. This indicates
that a QKD protocol is possible by using a n-particle GHZ state. 
Since a n-particle entangled state has little practical use, we will not
go into the details of describing the QKD protocol.

\section{Conclusion}

The quantum disentanglement eraser idea for 3-particle GHZ state has been
used here to construct two QKD protocols. The first one, where Alice holds
the 3-particle source, provides an additional level of security over the
BB84 or Eckert protocols. In ideal circumstances BB84 and Eckert methods provide
unbreakable key sharing, but in non-ideal cases several kinds of attacks
can be constructed. In such situations, our Master-Key Secured QKD protocol
will provide key-sharing which will be harder to break. We have also provided
a variant which we call Master-Key Controlled QKD where three
parties are involved. MKC-QKD allows the possibility for a third person,
called Master, to control the key sharing between Alice and Bob. Without
the {\em master-key} provided publicly by the Master at a later stage,
Alice and Bob will be unable to share a secure key. Various practical
uses of this method can be explored. For example, if the source of particles
is to be provided by a third party, this method can be used to establish
the authenticity of the source. This variant, however, is not expected to
provide and additional security over the BB84 or Eckert protocols.

\acknowledgments{T. Shibli thanks the Centre for Theoretical Physics for
the summer student program during which this work was completed. A. Sheel
thanks the Centre for Theoretical Physics for providing the facilities of
the Centre during the course of this work.}

\end{document}